\documentstyle[aps,prl,preprint]{revtex}
\begin{document}
\draft
\title{Low-temperature heat transfer in nanowires}\author{B.~A.~Glavin}
\address{Institute of Semiconductor Physics, Pr.~Nauki 45, Kiev 03028, Ukraine}
\date{\today}
\maketitle
\pacs{PACS numbers: 66.70.+f, 44.10.+i, 44.05.+e}
\begin{abstract}
The new regime of low-temperature heat transfer in suspended nanowires is
predicted. It takes place when (i)~only ``acoustic'' phonon modes of the wire
are thermally populated and (ii)~phonons are subject to the effective elastic
scattering. Qualitatively, the main
peculiarities of heat transfer originate due to appearance of the flexural
modes with high density of states in the wire phonon spectrum. They give rise to
the $T^{1/2}$ temperature dependence of the wire thermal conductance. The
experimental situations where the new regime is likely to be detected are
discussed.
\end{abstract}

Most often, in dielectrics and semiconductors heat is conducted by means of
phonon transfer. At low temperatures it is controlled by the phonon scattering
at the surface of the sample, since the internal mechanisms of scattering due to
the lattice anharmonicity and intrinsic defects provide the phonon mean free path
far above the typical dimensions of the sample. In this case the phonon transfer
is similar to the gas flow in small pipes in the Knudsen regime and, strictly
speaking, it is impossible to introduce local Fourier law for determination of
the heat flux density \cite{berman}. However, for the samples in the shape of a
wire it is still possible to speak about the thermal conductance, $\sigma_{T}$,
that is defined as a ratio of the heat flux through the wire and the temperature
difference between the reservoirs connected by the wire. If phonon scattering at
the surface is diffusive, then $\sigma_{T} \sim T^3$. Decrease of temperature,
however, increases the characteristic phonon wavelength and reduces the
scattering probability at the surface. If the probability of the specular phonon
scattering at the surface is close to unity, then the phonon spectrum is
substantially modified. For ideal wires having constant cross-section it is
broken in a set of branches where the phonon frequency $\omega$ depends on the1D
wavevector $q$ directed along the wire axis \cite{landau-lifschitz,nishiguci}.
In this letter we consider heat transfer at extremely low temperatures. Speaking
about the low temperatures, we mean that only the phonon states of the
``acoustic'' branches with $\omega \rightarrow 0$ at $q \rightarrow 0$ are
thermally populated. In other words, we assume $T \ll \hbar\Delta\omega$, where
$\Delta\omega$ is the characteristic value of the frequency gap between the
adjacent phonon branches, $\Delta\omega \sim v/a$, where $v$ and $a$ are the
characteristic sound velocity and dimension of the wire cross-section,
respectively.
This condition corresponds to the effective reduction of the phonon momentum
space dimensionality. It can be realized in the modern suspended films and
wires, employed for
the fundamental studies of the phonon confinement and electron-phonon
interaction \cite{schuller,wybourne}, and as elements of ultrasensitive phonon
detectors \cite{roukes-yocto}. Normally, condition $T \ll \hbar \Delta \omega$
corresponds to subkelvin temperatures. So, for $a=100~nm$ and $v = 5000~m/s$,
$\hbar\Delta\omega$ corresponds to the temperature about $0.4K$.
If the wire structure is perfect enough, then phonons are transferred between
the reservoirs ballistically. Ballistic transfer in wires was studied
theoretically in Refs.\cite{rego,roukes-thcond,blencowe}, where as thermal
conductance, as phonon correlation characteristics have been analyzed. It was
found that in the ballistic regime each phonon branch can contribute to the
value of the thermal conductance no more than the fundamental quantum $\pi
T/6\hbar^4$, regardless of the phonon spectrum details. This prediction
was recently confirmed experimentally \cite{roukes-nature}. In this letter we
consider the opposite case, when phonons are multiply scattered elastically
while being transferred through the wire. Similar problem was analyzed
previously in Refs.\cite{maynard,falko} where the numerical calculations of the
phonon transmission through the imperfect wire have been performed. In these
works, however, the authors used simplified models of the wire phonon spectrum
and miss some qualitative features of heat transfer. We demonstrate in this
letter, that for the case of effective phonon scattering at low temperatures the
thermal conductance is proportional to $T^{1/2}$. This feature is qualitatively
different as from the ballistic phonon transfer, as from the
surface-scattering-controlled transfer of 3D phonons. We show that the main
physical peculiarity of the 1D
phonon transfer in wires is due to the special features of the wire
phonon spectrum in the low-frequency region. Namely, some of the ``acoustic''
branches have
quadratic dispersion, and, therefore, high density of states and small group
velocity. This substantially affects as phonon scattering probabilities, as the
partial contributions of the phonons of particular type to the heat flux.
Note that similar effects arising due to the high density of states of the
low-dimensional phonons for electron-phonon interaction in thin films have been
reported recently \cite{submitted}. In the experiments, the $\sigma_T \sim
T^{1/2}$
dependence should replace $\sigma_T \sim T$, characteristic for the ballistic
phonon transfer, with temperature increase. We show, that the temperature
interval where the considered regime of heat transfer can be detected grows
with increase of the phonon scattering and the length of the wire.

As it is well established \cite{nishiguci}, in free-standing wires there are four ``acoustic'' branches having the following dispersion at $qa \ll 1$:
\begin{equation}\label{eq1}
\omega_{d,t} = v_{d,t} q,~ \omega_{f1,f2} = v_{f1,f2} a q^2.\end{equation}
Here $d$, $t$, $f1$, and $f2$ label the dilatational, torsional, and two flexural
branches. The values of $v_d$, $v_t$, $v_{f1}$, and
$v_{f2}$ are of the order of the sound velocity. Their specific values are
determined by the elastic properties of the wire material and by the shape of
the wire cross-section. To describe the steady-state heat transfer we use the
standard kinetic equation for the phonon distribution function $f$:
\begin{equation}
\label{eq2}
\frac{\partial f_i}{\partial x} g_i = St_i \{f\},
\end{equation}
where $i=d,t,f1,f2$; $g_i$ is the phonon group velocity, $x$-axis is directed
along the wire, and $St\{f\}$ is the
integral of collisions. Here we consider only elastic scattering of the phonons
by the defects of the wire structure. These defects can be either due to the
intrinsic
imperfections of the crystal, say, natural isotopes, or can arise in the course
of the nanowire fabrication process. Note that in actual suspended
nanowires the latter 
source, in particular, surface roughness, is likely to provide the major
contribution to the phonon scattering. We do not take into account three-phonon 
inelastic scattering due to the crystal anharmonicity. Having evaluated the 
corresponding phonon mean free path, we have found that it far exceeds typical 
length of the wires, about tens of microns.
Under this approach for $St\{f\}$ we have:
\begin{equation}\label{eq3}
St_i\{f\} = \sum_{j,q'} \left( W_{q'q}^{(ji)} f_j(q')(1+f_i(q)) - W_{qq'}^{(ij)}
f_i(q)(1+f_j(q'))\right),
\end{equation}
where $W_{q'q}^{(ji)}$ is the probability of the phonon transition $\{q',j\}
\rightarrow \{q,i\}$. We
consider defects whose dimensions are much less than $a$ and typical phonon
wavelength which corresponds to the Rayleigh scattering regime. To obtain
$W_{qq'}^{(ij)}$, it is necessary to write down the perturbation of the elastic
energy caused by a defect and calculate the phonon scattering probability using
the Fermi golden rule. Using the general form of the elastic energy
density,\cite{landau-lifschitz}
the leading term at small $q$ for the scattering probability can be written down
as
\begin{equation}\label{eq4}
W_{qq'}^{(ij)} = \frac{2\pi}{L} w_{ij} \frac{q^2 q'^2}{\omega \omega'} \delta
(\omega - \omega').
\end{equation}
Here factors $q^2$, $q'^2$ come from the spatial derivatives of the lattice
displacements in the expression for the elastic energy density, while $\omega$
and $\omega'$ in the denominator come from the expressions for the operators of
the lattice displacements. In Eq.~(\ref{eq4}) $L$ is the length of the wire, the
factor $2\pi$ is introduced for
simplicity. The factors $w_{ij}$ characterize the effectiveness of the
scattering and depend on
the spatial distribution, concentration, and characteristics of the defects.
Note, that the form of Eq.~(\ref{eq4}) is provided by the defects of the elastic
constants only. Contribution due to the mass defects is proportional to
$\omega\omega'$. As in the case of bulk crystals, for the transitions between
the 1D phonon branches having linear dispersion, the elastic-constant and mass
defects provide scattering probabilities with identical dependeces on the phonon
wavevectors. If, however, $i$ or $j$ corresponds to the flexural phonons, for
small $q$ contribution of the mass defects is much less then that of the
elastic-constant defects. 

For the introduced form of the scattering probabilities we obtain
\begin{equation}\label{eq5}
f_i^{(o)} = - \tau_i g_i \frac{\partial f_i^{(e)}}{\partial x},\end{equation}%
where $f_i^{(o)}$ and $f_i^{(e)}$ are odd and even parts of the distribution
function and $\tau_i$ is effective scattering time of $i$-th mode:
\begin{equation}
\label{eq6}\frac{1}{\tau_i} = \sum_{q',j} W_{qq'}^{(ij)}.
\end{equation}
Straightforward calculations demonstrate that if $qa \ll 1$ then the main
contribution to the $1/\tau_i$ is due to the scattering where the final state
belongs to one of the flexural modes, which is the direct result of their high
density of states:
\begin{eqnarray}\label{eq7}
\frac{1}{\tau_{f1}} = \left( \frac{w_{f1f2}}{v_{f2}^{3/2}} +
\frac{w_{f1f1}}{v_{f1}^{3/2}}\right) \frac{1}{q a^3 v_{f1}^{3/2}}, \nonumber \\
\frac{1}{\tau_{f2}} = \left( \frac{w_{f2f1}}{v_{f1}^{3/2}} +
\frac{w_{f2f2}}{v_{f2}^{3/2}}\right) \frac{1}{q a^3 v_{f2}^{3/2}}, \\
\frac{1}{\tau_{d}} = \left(\frac{w_{df1}}{v_{f1}^{3/2}}
+\frac{w_{df2}}{v_{f2}^{3/2}}\right) \frac{q^{1/2}}{v_d^{3/2} a^{3/2}},
\nonumber \\
\frac{1}{\tau_{t}} = \left(\frac{w_{tf1}}{v_{f1}^{3/2}}
+\frac{w_{tf2}}{v_{f2}^{3/2}}\right) \frac{q^{1/2}}{v_t^{3/2} a^{3/2}}.\nonumber
\end{eqnarray}
As we see, $\tau_{f1,f2} \rightarrow 0$ as $q \rightarrow 0$. This means that
semiclassical treatment of phonons by means of kinetic equation is invalid for
very small $q$. Introducing a characteristic frequency $\omega^\ast$ according
to $\tau_{f} |_{\omega = \omega^\ast} = (\omega^\ast)^{-1}$, we restrict
ourselves by consideration of regimes where $\hbar\omega^\ast \ll T$. In this
case, the major contribution to the heat flux is provided by the well-defined
phonon modes whose frequency exceeds considerably the scattering rate. Note, 
that similar wavevector dependence of scattering rate, as for flexural phonons, 
is inherent for scattering of electrons on short-range fluctuations of potential 
energy in 1D case. 

To determine the value of thermal conductivity we assume that $f_i^{(e)}$ have
Planck form with smooth spatial variation of temperature and obtain the
following expressions for the heat fluxes due to the phonons of each type, $j_i$:
\begin{eqnarray}\label{eq8}
j_{f1} = -8\zeta (3) \frac{1}{\hbar^2} T^2 a^3 v_{f1}^{3/2} \left(
\frac{w_{f1f2}}{v_{f2}^{3/2}} + \frac{w_{f1f1}}{v_{f1}^{3/2}}\right)^{-1}
\frac{dT}{dx},  \nonumber\\
j_{f2} = -8\zeta (3) \frac{1}{\hbar^2} T^2 a^3 v_{f2}^{3/2} \left(
\frac{w_{f2f1}}{v_{f1}^{3/2}} + \frac{w_{f2f2}}{v_{f2}^{3/2}}\right)^{-1}
\frac{dT}{dx}, \\
j_d = -\pi^{1/2} \zeta (3/2) v_d^3 a^{3/2} \left(\frac{T}{\hbar}\right)^{1/2}
\left( \frac{w_{df1}}{v_{f1}^{3/2}} + \frac{w_{df2}}{v_{f2}^{3/2}}\right)^{-1}
\frac{d T}{d x},\nonumber \\
j_t = -\pi^{1/2} \zeta (3/2) v_t^3 a^{3/2}  \left(\frac{T}{\hbar}\right)^{1/2}
\left( \frac{w_{tf1}}{v_{f1}^{3/2}} + \frac{w_{tf2}}{v_{f2}^{3/2}}\right)^{-1}
\frac{d T}{d x}, \nonumber
\end{eqnarray}
where $\zeta$ stands for Riemann function. It is easy to see that at $T\ll \hbar
\Delta \omega$ the dilatational and torsional phonons provide major contribution
to the heat flux, and, therefore, to the thermal conductivity and thermal
conductance $\sigma_T = \kappa/L$:
\begin{equation}\label{eq9}
\kappa = \pi^{1/2} \zeta (3/2) \left(\frac{T}{\hbar}\right)^{1/2} a^{3/2} \left(
v_d^3 \left( \frac{w_{df1}}{v_{f1}^{3/2}} +
\frac{w_{df2}}{v_{f2}^{3/2}}\right)^{-1} + v_t^3 \left(
\frac{w_{tf1}}{v_{f1}^{3/2}} + \frac{w_{tf2}}{v_{f2}^{3/2}}\right)^{-1} \right).
\end{equation}
From Eqs.~(\ref{eq7},\ref{eq8}) we can finally elucidate the role of the
flexural phonons in heat transfer. First, their contribution to the heat
flux is small due to their slowness and high scattering probability. Indeed,
according to Eq.~(\ref{eq4}) the latter is proportional to the square of the
initial phonon wavevector. If we consider phonons having energy about $T$, then
for the flexural phonon the value of $q^2$ is greater than for the dilatational
and torsional phonons roughly with the factor $\hbar \Delta \omega /T$. Second,
as a result of their high density of states, the flexural phonons provide very
effective scattering for the dilatational and torsional phonons. It is worth to
mention that the flexural phonons are important also for the nonstationary heat
transfer. The latter is characterized by both thermal conductivity and thermal
capacity per unit length, $c$. It can be easily obtained that for $T \ll \hbar
\Delta \omega$ $c \sim T^{1/2}$ and the flexural phonons provide major
contribution to $c$, again, due to their high density of states.

An important point that is worth to be discussed is about the influence of the
phonon localization on the heat transfer. According to the general predictions
of the localization theory, if only elastic scattering is present and there is
no source of dephasing, the conductance of a 1D system decays exponentially with
its length $L$ if $L$ exceeds the mean free path of a carrier. We believe,
however, that the described regime of heat transfer is not cancelled by
localization, in general. This is because we deal with the multichannel transfer
regime where the role of distinct channels is essentially different. In
particular, the main contribution to the heat flux is due to the dilatational
and torsional phonons, while the flexural phonons provide major contribution to
the overall phonon density of states. Using semi-quantitative arguments of
Thouless \cite{thouless}, we expect that localization is likely to be manifested
for $L > l (\hbar\Delta\omega/T)^{1/2} \gg l$, where $l$ is the characteristic
mean free path of the dilatational and torsional phonons. Nevertheless, we would
like to stress that the problem of localization definitely deserves more
rigorous consideration. As we demonstrated, the peculiarities of the phonon
spectrum of a wire bring about asymptotic behavior of the phonon mean free path
which is qualitatively different from that of bulk crystals. This can cause
qualitatively new features of localization phenomena.

Finally, it is necessary to analyze the experimental situations where the
described regime of heat transfer can be detected. This directly follows from
the main assumptions we made. First, the condition $T\ll \hbar \Delta\omega$
must be satisfied. Second, phonons must be multiply scattered during the
transfer through the wire. The first condition depends on the wire cross-section
and material parameters only, while the second one depends on the wire length
and characteristics of defects. Since characteristic phonon mean free path
decreases as the temperature increases, the temperature range where the
predicted heat transfer regime can be detected expands for the longer and less
perfect wires. This is qualitatively illustrated in Fig.~1, where we plot
schematically the temperature dependence of the wire thermal conductance. The
vertical line marks the temperature where the condition $T = \hbar \Delta
\omega$ is reached. The dashed line corresponds to the ballistic phonon
transfer, $\sigma_T \sim T$, while the two dotted curves describe the case of
intensive phonon scattering, considered in this letter, $\sigma_T \sim T^{1/2}$.
Obviously, if the wire cross-section is fixed, the upper dotted curve
corresponds to the wire which is either shorter or more perfect, since $\sigma_T
\sim 1/(Lw)$. The two solid lines show the resulting temperature dependence of
the thermal conductance. We see that in the wire that is longer or less perfect
the $T^{1/2}$ law is realized in the wider temperature range and is likely to be
detected. Note, that some additional features, not shown in Fig.~1, can appear
at temperatures below the $T$ to $T^{1/2}$ transition. They arise because
the flexural phonons are scattered much more effectively than the dilatational
and torsional phonons. Therefore, at some intermediate temperatures only the
dilatational and torsional phonons are ballistic. In this case $\sigma_T \sim
T$, but the value of the coefficient is twice lower, than for the case where
all ``acoustic'' phonons are ballistic.
In experiment \cite{roukes-nature} sublinear temperature dependence of the
wire thermal conductance was detected in some temperature range. However, this
feature can appear not only due to the phonon scattering described in this
letter, but also due to the imperfect acoustic coupling between the wire and the
reservoirs. It is necessary to undertake additional studies to determine which
of these two reasons is responsible for the observed behavior.

In conclusion, we have predicted the new regime of heat transfer in nanowires at low
temperatures, where only the lowest branches of the wire phonon spectrum are
thermally populated. It has been demonstrated that the flexural phonons having
quadratic dispersion play especial role. They accumulate most of the wire
thermal energy, practically do not contribute to the heat flux, and, finally,
provide effective scattering to the dilatational and torsional phonons, that
carry most of the heat flux. We have found that in this case thermal conductivity is
proportional to $T^{1/2}$. Experimentally, this heat transfer regime is most
likely to be detected in the samples with relatively high phonon scattering.

The author would like to thank Prof.~V.~A.~Kochelap for fruitful
discussions. This work was supported partially by the ERO of US Army under 
Contract~No.~N68171-01-M-5166, fellowship of the Institute of Semiconductors, 
Frankfurt (Oder), and the Russia-Ukraine Program for Nanoelectronics.

\begin{figure}
\caption{
Schematic temperature dependence of the wire thermal conductance in the 
different
transport regimes. Dashed curve corresponds to the ballistic phonon transfer,
while the two dotted lines correspond to the phonon transfer with intensive
scattering, and the solid line represent the resulting temperature dependence of
$\sigma_T$. Vertical line marks the temperature where the upper phonon branches
become thermally populated. The upper dotted line represents the dependence for
the wire which is shorter or is characterized by less intensive phonon
scattering.
}
\end{figure}
\end{document}